\documentclass[pra,aps,twocolumn,showpacs]{revtex4}
\usepackage{bm}
\usepackage{amsmath,amssymb}
\usepackage{graphicx}

\begin{document}

\title{Xenon Clusters in Intense VUV Laser Fields}
\author{Robin Santra}
\author{Chris H. Greene}
\affiliation{Department of Physics and JILA, University of Colorado, Boulder, CO 80309-0440, USA}
\date{\today}
\begin{abstract}
A simple model is developed that quantitatively describes intense interactions of a VUV laser 
pulse with a xenon cluster. We find good agreement with a recent experiment [H. Wabnitz {\em et al.}, Nature 
{\bf 420}, 482 (2002)]. In particular, the large number of VUV photons absorbed per atom---at 
intensities significantly below $10^{16}$~W/cm$^{2}$---is now understood.
\end{abstract}
\maketitle

Little is known about laser--cluster interactions at UV or higher photon energies. The
destructive impact of laser pulses with a peak intensity of almost $10^{19}$~W/cm$^{2}$
at a wavelength of $248$~nm was demonstrated by McPherson {\em et al.} \cite{McTh94}.
However, intense laser fields at even higher photon energies have not been accessible until
very recently. In 2000, the first lasing---in a free-electron laser (FEL)---at
$\lambda = 109$~nm was reported \cite{AnAu00}. The FEL is part of the TESLA Test Facility
(TTF) in Hamburg, Germany. One of the major objectives of the TTF is the development of the
technology for an ultrabright x-ray laser \cite{MaTs01}. The new VUV laser source has already 
displayed its capability for exploring interesting physics: Motivated by the outstanding 
properties of the radiation generated by the TTF FEL, documented in Ref.~\cite{AyBa02}, 
researchers in Hamburg exposed xenon atoms and clusters to the intense VUV laser pulses 
\cite{WaBi02}.

Each laser pulse had a duration of about $100$~fs and consisted of $12.7$-eV photons. The
highest intensity in the experiment was about $7 \times 10^{13}$~W/cm$^{2}$ \cite{MoPr03}. Under these
conditions, {\em isolated} Xe atoms are found to be only singly ionized. This observation is
compatible with one-photon absorption, as the ionization threshold of Xe is $12.1$~eV
\cite{BrVe01}. Multiphoton processes apparently are of no relevance. Clusters of $1000$
atoms or more, on the other hand, behave in a strikingly different way: They absorb at least
$30$ VUV photons per atom. The clusters are completely destroyed, and ion charge states of
up to $8+$ can be detected.

It is the purpose of this Letter to elucidate the physics underlying the experimental observations. 
We show that the high efficiency of VUV photon absorption in xenon clusters is due to inverse bremsstrahlung  
\cite{SeHa73,KrWa73,ShYa75} in combination with atomic-structure and plasma-screening effects. 
Nonlinear optical processes do not play a role.

Considering the relatively low intensity of the TTF FEL, the experimental findings are rather 
surprising. Producing similarly pronounced ionization and fragmentation phenomena in noble gas 
clusters using near-infrared lasers requires pulse intensities of $10^{16}$~W/cm$^{2}$ or higher
\cite{HuDi98}. Under these circumstances, the clusters are turned into microplasmas, and x-ray 
emission \cite{DiDo95} and highly energetic electrons \cite{ShDi96} and ions \cite{LeDo98} can 
be observed. 

The high intensity at long wavelengths serves two purposes \cite{DiDo96,RoSc97,LaJo99,SiRo02}.
On one hand, even though a noble gas atom cannot be ionized by single near-infrared photons, some 
of the valence electrons can tunnel through, or even escape over the potential barrier generated 
by the ionic core and the strong quasistatic electric field of the laser. On the other hand, the 
average kinetic energy, or ponderomotive potential, of an electron oscillating in the laser field 
can easily be of the order of $1$~keV. This energy can be released in collisional ionization, i.e., 
($e,2e$) type reactions. Moreover, in energetic electron--ion collisions a substantial number of 
photons can be absorbed from the laser field. This heating mechanism is referred to as inverse 
bremsstrahlung (IBS). 

At VUV photon energies, however, the ponderomotive potential is only on the order of $10$~meV. 
Indeed, numerical simulations and estimates in Ref.~\cite{WaBi02}, which are based on methods 
developed in the context of low-frequency lasers \cite{DiDo96,LaJo99}, predict the absorption 
of only a few photons per atom. This differs from the experimental result by more than an order 
of magnitude. 

\begin{figure}[h]
\includegraphics[width=6.5cm,origin=c,angle=-90]{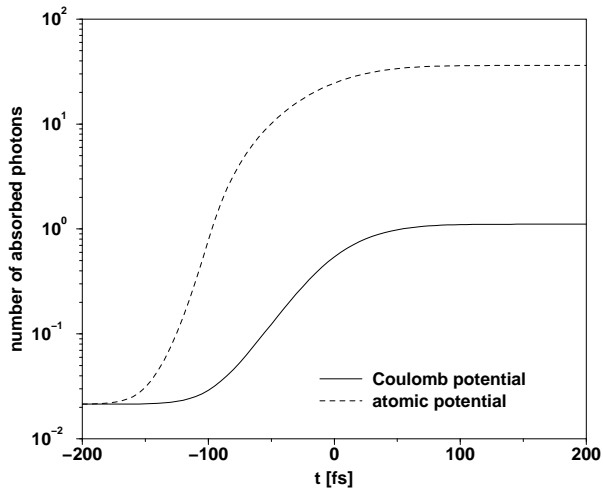}
\caption[]{Number of VUV photons ($\hbar \omega$ = $12.7$~eV) absorbed per atom via
inverse bremsstrahlung, plotted as a function of time. All atoms within the cluster are
assumed to be singly ionized prior to the laser pulse. The average initial kinetic energy
of a plasma electron is $0.01$~hartree. The laser pulse, having a peak intensity of
$7.3 \times 10^{13}$~W/cm$^{2}$, is Gaussian. It is centered at $t=0$; its FWHM is $100$~fs.
If the electrons are scattered by simple (Debye-screened) Coulomb potentials, only one
photon per atom is absorbed from the laser field. About $30$ photons are absorbed, however, if a
more realistic atomic scattering potential is taken.}
\label{fig1}
\end{figure}

The photoionization cross section of neutral Xe at $\hbar\omega = 12.7$~eV is roughly $50$~Mb \cite{Sams66}.
Hence, a pulse with an intensity of $10^{13}$~W/cm$^{2}$ ionizes all atoms in a cluster within
$10$~fs. The resulting plasma bears some similarity to a metal: The plasma electrons can move
freely, but because of the high atomic density there is a high probability for electron--ion
and electron--electron collisions. Free electrons cannot absorb photons; neither can photons be
absorbed in electron--electron collisions \cite{ShYa75}.

However, electron--ion collisions can extract energy from the laser field via IBS. Since we are
considering relatively moderate intensities and rather short wavelengths, it is legitimate
to treat this process perturbatively. Using second-order perturbation theory (first order
in electron--ion and electron--photon coupling, respectively) and assuming a cluster of infinite
spatial extension but constant atomic density, one can derive a quantum-mechanical formula for 
the heating rate per plasma electron. In our implementation of the IBS heating rate, we avoided 
making use of the classical limit $\hbar \rightarrow 0$, which is often taken in the weak-field 
case (see, for example, Ref.~\cite{SeHa73}), but which is inappropriate for VUV photons. We 
exploit, however, the fact that due to numerous collisions with ions and electrons all 
directionality imprinted on the plasma electrons by the linear polarization of the VUV laser 
field is lost and thermal equilibrium among the plasma electrons is maintained. Finally, the plasma 
electrons are assumed to form a nondegenerate gas, and at each point in time all electrons experience 
the same laser electric field. Taking the thermodynamic limit and simultaneously utilizing the 
electric-dipole approximation for the entire system is reasonable for a cluster with a radius of 
$100$~{\AA} or so, comprising many thousands of atoms.

Let all atoms in the cluster be singly ionized (we treat the combined effect of photoionization
and IBS later), and let the atomic density be that of liquid xenon. To a first approximation, 
the plasma electrons are scattered by point-like ions of charge $+1$. Additionally, the plasma 
electrons can screen the ionic field. Using a Debye-shielded Coulomb potential \cite{Stur94}, 
we calculated the number of VUV photons absorbed per atom via IBS, a plasma electron having an 
average initial kinetic energy of $0.01$~hartree. (Screening due to ions as well as ionic motion 
during the laser pulse are neglected throughout.) The result, as a function of time, is shown in 
Fig.~\ref{fig1}. Our data are based on a Gaussian laser pulse, centered at $t=0$ with a FWHM of 
$100$~fs. The peak intensity is $7.3 \times 10^{13}$~W/cm$^{2}$. After the pulse is over, each 
plasma electron has absorbed only a single photon, which is clearly in disagreement with 
experiment, but analogous to the estimate quoted in Ref.~\cite{WaBi02}.

The problem is the ionic scattering potential. In the dense plasma, the electrons experience
more than just a simple Coulomb field. A more realistic treatment of the atomic potential is
needed. We use the form
\begin{equation}
\label{eq1}
V_i(r) = -\frac{1}{r}\left\{i + [Z - i] \exp{(-\alpha_i r)}\right\}\exp{(-r/\lambda_D)},
\end{equation}
where $i$ is the ionic charge state, $Z = 54$ the nuclear charge, and $\lambda_D$ the Debye
length. The parameter $\alpha_i$ controls the transition from the exterior of the ion to 
its interior, where a colliding electron experiences an effective charge higher than $i$.
We adjust $\alpha_1$ in such a way that the binding energy of a $5p$ electron in the 
potential $V_1(r)$ (for $\lambda_D \rightarrow \infty$) equals the ionization potential of
neutral Xe. That the resulting potential is useful for quantitative predictions can be 
illustrated by calculating the photoionization cross section: At $12.7$~eV, we find agreement
with experiment to within $10$\%. (Both the binding energy and the photoionization cross section
are calculated by numerically diagonalizing the one-electron Hamiltonian based on $V_1(r)$.)
If we now suppose that a plasma electron scattered by Xe$^+$ experiences the same potential, 
$V_1(r)$, then the number of absorbed VUV photons per atom turns out to be about $30$. This is 
illustrated in Fig.~\ref{fig1}. The mechanism of IBS is, thus, indeed capable of explaining the 
huge amount of VUV laser energy deposited in a large xenon cluster.

\begin{figure}[h]
\includegraphics[width=6.5cm,origin=c,angle=-90]{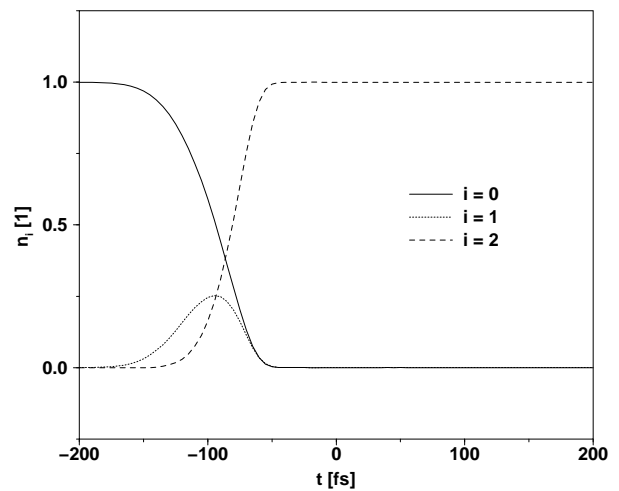}
\caption[]{Temporal evolution of neutral and ionic populations in a large xenon
cluster exposed to an intense VUV laser pulse. The probability of finding a neutral
atom in the cluster is $n_0$; $n_1$ and $n_2$ refer to singly and doubly ionized species,
respectively. The same laser parameters as in Fig.~\ref{fig1} are used. Plasma screening,
which is responsible for efficient double photoionization of xenon, is taken into account.
Collisional ionization is not included.}
\label{fig2}
\end{figure}

So far we have restricted ourselves to a preformed plasma interacting with a laser pulse.
In order to arrive at a more complete picture, we need to follow the time evolution of 
photoionization and collisional heating. Since we anticipate plasma screening effects to lower
the ionization thresholds, our treatment is not restricted to single photoionization. We 
formulate a set of coupled rate equations for the time-dependent probabilities $n_i(t)$ 
($i \ge 0$) of finding Xe$^{i+}$ in the cluster: 
\begin{eqnarray}
\label{eq2}
\dot{n}_0(t) & = & -\sigma_1(t) j_{\mathrm{ph}}(t) n_0(t) \nonumber \\
\dot{n}_1(t) & = & \sigma_1(t) j_{\mathrm{ph}}(t) n_0(t) - \sigma_2(t) j_{\mathrm{ph}}(t) n_1(t) \\
& \vdots & \nonumber 
\end{eqnarray}
Here, $j_{\mathrm{ph}}(t)$ is the photon flux. These rate equations require knowledge of the 
photoionization cross sections $\sigma_{i+1}(t)$ of the Xe$^{i+}$ species embedded in the Debye plasma. 
Taking the double \cite{HaPe87}, triple \cite{MaBa87}, and quadruple \cite{GrDi83} ionization thresholds 
from the literature, we can determine, in addition and in analogy to $\alpha_1$, the parameters 
$\alpha_2$, $\alpha_3$, and $\alpha_4$ (see Eq.~(\ref{eq1})). (In so doing, we imply that the various 
$5p$ levels have---at least approximately---the same energy.) With this information, the photoionization 
cross sections of Xe, Xe$^{+}$, Xe$^{++}$, and Xe$^{3+}$ can be calculated as a function of time. 
Note that the Debye length in Eq.~(\ref{eq1}) is time-dependent, for it is a function of the 
temperature and density of the plasma electrons. 

The rate equations governing the populations $n_i(t)$ are complemented by a rate equation that
describes heating of the plasma electrons:
\begin{equation}
\label{eq3}
\dot{{\cal E}}_{\mathrm{kin}}(t) = \frac{3}{2} q(t) \dot{T}(t) + \sum_{i} \varepsilon_i(t)
\sigma_i(t) j_{\mathrm{ph}}(t) n_{i-1}(t),
\end{equation}
where $q(t)$ is the average number of plasma electrons per atom, T(t) the electron temperature (in units
of energy), and ${\cal E}_{\mathrm{kin}}(t) = \frac{3}{2} q(t) T(t)$. $\varepsilon_i(t)$ denotes the kinetic 
energy of a photoelectron leaving Xe$^{i+}$ behind. Two contributions are taken into account in 
Eq.~(\ref{eq3}): the collisional heating rate due to IBS ($\dot{T}(t)$) and the kinetic energy of 
photoelectrons newly added to the plasma. This rate equation also depends on the neutral and ionic 
populations as well as the respective photoionization cross sections.

\begin{figure}[h]
\includegraphics[width=6.5cm,origin=c,angle=-90]{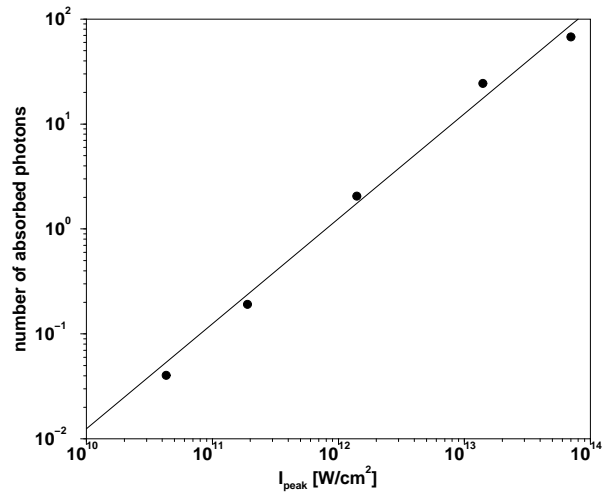}
\caption[]{Number of VUV photons absorbed per atom via inverse bremsstrahlung, plotted as a function of
peak pulse intensity. In the underlying calculations, coupled rate equations for the neutral and ionic
populations as well as for the temperature of the plasma electrons are numerically integrated.
Initially, all atoms are in their neutral ground state. The range of intensities shown was explored in
the recent cluster experiment at the TESLA Test Facility in Hamburg \cite{WaBi02}. The solid line is a
linear fit to the numerical data.}
\label{fig3}
\end{figure}

Starting with a cluster of neutral atoms in their ground state, we numerically integrate the entire 
set of coupled rate equations (Eqs.~(\ref{eq2}) and (\ref{eq3})). For laser pulse parameters identical 
to the ones employed for Fig.~\ref{fig1}, the probabilities $n_i(t)$ shown in Fig.~\ref{fig2} are obtained. 
As expected, the neutral population is completely depleted after just $10$~fs or so. It is more interesting
to observe that singly ionized xenon also vanishes on the same timescale. Due to plasma screening,
the production of Xe$^{++}$ becomes energetically accessible. Higher ionic charge states, however,
cannot be generated by direct photoionization.

According to Fig.~\ref{fig2}, all xenon atoms are doubly ionized before the laser pulse even reaches
its maximum. Therefore, each atom contributes two electrons to the plasma, which results in 
enhanced energy absorption: Almost $70$ photons per atom are taken from the VUV laser field at a peak 
intensity of $7.3 \times 10^{13}$~W/cm$^{2}$, as can be seen in Fig.~\ref{fig3}. Also shown in that 
figure are the corresponding data for lower laser intensities. Clearly, reduced intensities lead to a 
smaller number of plasma electrons and to less collisional heating. On the log--log scale of 
Fig.~\ref{fig3}, the effect may appear to be dramatic. However, the relationship between the number of 
absorbed photons and the peak intensity of the laser pulse is really a linear one. The solid line in 
Fig.~\ref{fig3} represents a linear fit to the data.

There is no optical nonlinearity involved. In fact, our model of photoionization and IBS does not 
contain true multiphoton physics. Nevertheless, at high intensity, many photons are absorbed by
each plasma electron. But this does not happen in a single step. Each plasma electron is scattered
many times and can absorb only a single photon during a collision with an ion.

\begin{figure}[h]
\includegraphics[width=6.5cm,origin=c,angle=-90]{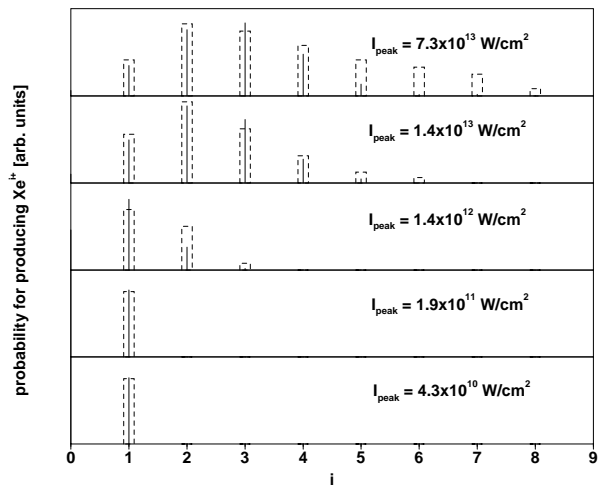}
\caption[]{Ionic populations resulting from the interaction of a $100$-fs long VUV laser pulse
with a large xenon cluster. The solid lines are obtained within a simple rethermalization model
(please see the text). The dashed bars reflect mass-spectroscopic signal strengths observed for
an average cluster size of $1500$ atoms, estimated from data in Ref.~\cite{WaBi02}.}
\label{fig4}
\end{figure}

Up to this point, we have ignored collisional ionization. This effect can be incorporated 
{\em a posteriori} by noticing that electron--ion collisions facilitate a thermal equilibrium 
among the plasma electrons and the ions. Thus, the electron gas is cooled, and energy is 
transferred to the ions. In order to quantify this picture, the first eight ionization potentials of 
xenon are needed. To this end, we made use of the complete-active-space self-consistent-field code 
implemented in the {\em ab initio} package MOLPRO \cite{WeKn85,KnWe85} and of the effective core 
potential by LaJohn {\em et al.} \cite{JoCh87}. Our first four ionization potentials are in agreement 
with experiment \cite{BrVe01,HaPe87,MaBa87,GrDi83} at a level of a few percent. We assign a similar 
accuracy to our calculation of the energies required to turn a neutral xenon atom into even
more highly charged ions: $156$~eV for Xe$^{5+}$, $220$~eV for Xe$^{6+}$, $310$~eV for Xe$^{7+}$,
and $414$~eV for Xe$^{8+}$. (It is justified to use unscreened ionization potentials at the end of
the laser pulse, in view of the fact that at high intensities the electron kinetic energies are rather
high and at low intensities there are just a few electrons that could contribute to shielding.)

From the solution of our coupled rate equations, we determine the total laser energy stored in the
electrons and ions. This energy is then redistributed assuming: a Boltzmann distribution for the
ionic charge states; thermal kinetic energies for electrons and ions; the existence of a common
temperature. A statistical weight is introduced for each ionic charge state by simply counting the
number of ways the valence electrons can be distributed over the $5p$ spin orbitals (Xe, ..., Xe$^{5+}$;
the $5s$ level being doubly occupied) or over the $5s$ spin orbitals (Xe$^{6+}$, Xe$^{7+}$). A single state
is assigned to Xe$^{8+}$. (Other electronic states of the Xe$^{i+}$ are neglected.) The ionic populations 
resulting from this procedure are plotted in Fig.~\ref{fig4}, together with experimental data taken from 
Ref.~\cite{WaBi02}. Considering the simplicity of our approach, it is amazing how well the experimental 
Xe$^{i+}$ populations for various laser intensities can be reproduced. We learn from Fig.~\ref{fig4}
that the experimental signal, obtained after fragmentation of the clusters, basically reflects the thermal 
ionic distribution in the plasma.

In this Letter, we have demonstrated that some of the concepts familiar from low-frequency laser--cluster 
physics can be transferred to higher photon energies. However, this transfer requires a more detailed 
description of atoms in plasmas than previously anticipated. Similar care is therefore necessary when 
evaluating the potential of future x-ray lasers for imaging of single molecules \cite{NeWo00}. If not all 
physically relevant processes are taken into account, for example interatomic electron-correlation phenomena 
following core-hole relaxation \cite{SaCe03}, then the degree of damage caused in a large biomolecule, for 
instance, can be easily underestimated. 

We believe that the theoretical description presented here provides insight into the nature of intense 
laser--cluster interactions at VUV wavelengths. Yet we anticipate that, in this new laser regime of
higher photon energies, many more surprises are likely to emerge.

\acknowledgments
R.S. gratefully acknowledges financial support by the Emmy Noether program of the 
German Research Foundation (DFG). This work was supported in part by the Department 
of Energy, Office of Science.


\begin{thebibliography}{99}

\bibitem{McTh94} A. McPherson {\em et al.}, Nature {\bf 370}, 631 (1994).

\bibitem{AnAu00} J. Andruszkow {\em et al.}, Phys. Rev. Lett. {\bf 85}, 3825 (2000).

\bibitem{MaTs01} {\em TESLA Technical Design Report: The X-Ray Free Electron Laser}, 
edited by G. Materlik and Th. Tschentscher (DESY, Hamburg, 2001), Vol. V.

\bibitem{AyBa02} V. Ayvazyan {\em et al.}, Phys. Rev. Lett. {\bf 88}, 104802 (2002).

\bibitem{WaBi02} H. Wabnitz {\em et al.}, Nature {\bf 420}, 482 (2002).

\bibitem{MoPr03} According to a more recent analysis, the maximum intensity was 
$3 \times 10^{13}$~W/cm$^{2}$ at a pulse duration of about $50$~fs
(T. M\"{o}ller, private communication).

\bibitem{BrVe01} F. Brandi {\em et al.}, Phys. Rev. A {\bf 64}, 032505 (2001).

\bibitem{SeHa73} J. F. Seely and E. G. Harris, Phys. Rev. A {\bf 7}, 1064 (1973).

\bibitem{KrWa73} N. M. Kroll and K. M. Watson, Phys. Rev. A {\bf 8}, 804 (1973).

\bibitem{ShYa75} Y. Shima and H. Yatom, Phys. Rev. A {\bf 12}, 2106 (1975).

\bibitem{HuDi98} M. H. R. Hutchinson {\em et al.}, Phil. Trans. R. Soc. Lond. A
{\bf 356}, 297 (1998).

\bibitem{DiDo95} T. Ditmire {\em et al.}, Phys. Rev. Lett. {\bf 75}, 3122 (1995).

\bibitem{ShDi96} Y. L. Shao {\em et al.}, Phys. Rev. Lett. {\bf 77}, 3343 (1996).

\bibitem{LeDo98} M. Lezius {\em et al.}, Phys. Rev. Lett. {\bf 80}, 261 (1998).

\bibitem{DiDo96} T. Ditmire {\em et al.}, Phys. Rev. A {\bf 53}, 3379 (1996).

\bibitem{RoSc97} C. Rose-Petruck {\em et al.}, Phys. Rev. A {\bf 55}, 1182 (1997).

\bibitem{LaJo99} I. Last and J. Jortner, Phys. Rev. A {\bf 60}, 2215 (1999).

\bibitem{SiRo02} C. Siedschlag and J. M. Rost, Phys. Rev. Lett. {\bf 89}, 173401 (2002).

\bibitem{Sams66} J. A. R. Samson, in {\em Advances in Atomic and Molecular Physics}, 
edited by D. R. Bates and I. Estermann (Academic Press, New York, 1966), Vol.~2.

\bibitem{Stur94} P. A. Sturrock, {\em Plasma Physics} (Cambridge University Press, 
Cambridge, 1994).

\bibitem{HaPe87} J. E. Hansen and W. Persson, Phys. Scr. {\bf 36}, 602 (1987).

\bibitem{MaBa87} D. Mathur and C. Badrinathan, Phys. Rev. A {\bf 35}, 1033 (1987).

\bibitem{GrDi83} D. C. Gregory, P. F. Dittner, and D. H. Crandall, Phys. Rev. A {\bf 27}, 
724 (1983).

\bibitem{WeKn85} H.-J. Werner and P. J. Knowles, J. Chem. Phys. {\bf 82}, 5053 (1985).

\bibitem{KnWe85} P. J. Knowles and H.-J. Werner, Chem. Phys. Lett. {\bf 115}, 259 (1985).

\bibitem{JoCh87} L. A. LaJohn {\em et al.}, J. Chem. Phys. {\bf 87}, 2812 (1987)

\bibitem{NeWo00} R. Neutze {\em et al.}, Nature {\bf 406}, 752 (2000).

\bibitem{SaCe03} R. Santra and L. S. Cederbaum, Phys. Rev. Lett. {\bf 90}, 153401 (2003).

\end{thebibliography}
\end{document}